\documentclass[times, 10pt,twocolumn]{article}

\usepackage{times}
\usepackage{latex8}
\usepackage{cite}
\usepackage{graphicx}
\usepackage{url}

\usepackage{url} 
\usepackage{graphicx}
\usepackage{latexsym}
\usepackage{cite}
\usepackage{balance}

\newcommand{\pf}{ProdFLOW$^{\textrm{\small{TM}}}$}

\begin{document}

\title{A Systematic Review of Productivity Factors in Software Development}

\author{Stefan Wagner\\
Technische Universit\"at M\"unchen\\
Garching b.~M\"unchen, Germany\\
wagnerst@in.tum.de
\and 
Melanie Ruhe\\
Siemens AG\\
Munich, Germany\\
melanie.ruhe@siemens.com
}

\maketitle
\thispagestyle{empty}

\begin{abstract}
Analysing and improving productivity has been one of the main goals
of software engineering research since its beginnings. A plethora 
of studies has been conducted
on various factors that resulted in several models for analysis and prediction
of productivity. However, productivity is still an issue in current
software development and not all factors and their relationships are known.
This paper reviews the large body of available literature in order to
distill a list of the main factors influencing productivity investigated
so far. The measure for importance here is the number of articles a factor
is mentioned in. Special consideration is given to \emph{soft} or human-related
factors in software engineering that are often not analysed with equal
detail as more technical factors. The resulting list can be used to guide
further analysis and as basis for building productivity models.
\end{abstract}

\Section{Introduction}

Productivity in software development has been an important research area
for several decades. It is the key for a successful software company to
control and improve its productivity. However, in contrast to traditional
industrial work, it is hard to measure for software development. There
are several terms that are used more or less synonymously such as
performance or efficiency. There are also various definitions
of which output divided by input is the most general one.

In software development lines-of-code (LOC) and function points (FP) are
traditionally used in measures for productivity, i.e., the amount of LOC or FP
produced per hour by a developer. Based on this, there is a large amount
 of studies on various aspects of productivity. The two mentioned measures
and several more dimensions have been analysed and detailed. Models have
been built that should explain, analyse and predict productivity.
Finally, several studies analyse the factors that influence productivity
in a software project.

\paragraph{Problem}
Frese and Brodbeck \cite{frese94} claim that the scientific discussion about
the work situation in software development and about productivity
factors in such projects is done based on an insufficient empirical basis. 
According to them, it
is dominated by shallow surveys and qualitative experience reports.

Moreover, the software engineering literature in that area often has a
strong emphasis on mainly technical factors such as the software size or
the product complexity. However, Brodbeck \cite{brodbeck94a} shows that 
more than a third of the time a typical software developer is not 
concerned with technical work.
Meetings and talks constitute 21.1\%, presentations and project organisation
9.6\%, and independent qualification 6\% of the work time. Hence, these
efforts are significant.

\paragraph{Contribution}

We provide a systematic review of software engineering, management 
and organisational psychology literature on productivity
factors in software development. A list of these factors is distilled
from the literature in order to aid model building, productivity
improvement and further research.

The importance of the factors and thereby their inclusion in the
list is based on their mentioning in the analysed literature. Hence,
it is secured that several authors used and/or analysed the factor.
Furthermore, we put specific care to the equal consideration of 
technical and soft factors in order to represent their significance.

%

\Section{Review approach}
\label{sec:approach}

For the systematic review of the productivity literature, we use
a combined approach of automated and manual search. Our aim is
to include literature from the areas software engineering, 
management and organisational psychology as these are the main
sources of relevant literature. For
this we used a query on four portals for scientific literature
that contains the typically used terms in papers about what
we call \emph{productivity factors}.
For the automated search we used the following expression:
\begin{verbatim}
software AND (productivity OR "development 
efficiency" OR "development effectiveness" 
OR "development performance")
\end{verbatim}

It resulted in the following numbers of
results:
\begin{itemize}
\item ACM's The Guide: 10,017
\item IEEE Xplore: 1,408
\item ScienceDirect: 508
\item Google Scholar: 680,000
\end{itemize}

These large numbers show on the one hand the significance of the
topic in research but on the other hand prevents the manual analysis
of all these papers. We inspected the first 100 results of each portal
whether they are suitable for inclusion in our study. In this inspection
we omitted very specific analyses of single, detailed factors because
of brevity reasons. We also excluded studies that only showed that
factors have no influence on productivity. Although these studies are
of interest in general, they do not help in building a list of factors
that do influence productivity.

In addition to the papers retrieved using that query, we also collected
papers manually in a number of important journals in software
engineering (e.g.\ IEEE Transactions on Software Engineering),
in management (e.g.\ Management Science) and organisational
psychology (e.g.\ Journal of Occupational and Organizational
Psychology). From these and by following references from the already
found papers, the complete body of papers that build the basis of
this study was collected. Moreover, despite some limitations, the well-known
books by Boehm \cite{boehm88} and Jones \cite{jones00} on software productivity
were included as a baseline.

We derived from this body of papers factors about the product, process and
people and unified synonymous terms as far as possible. Then the
extracted factors were ranked by appearances in the literature. We mainly
aimed at finding different authors that used the factors. Based on
this, the final list was compiled.

\Section{Considered studies}
\label{sec:studies}

Because of space limitation we cannot describe each considered study
but we only choose some important representatives of each decade. The 
full description can be found in \cite{wagner:tr_survey}.
We decided to organise the
papers in the order of their publication which has the additional benefit
that the developments over time become visible. 

\SubSection{1970--1979}

Walston and Felix \cite{walston77} analysed in 1977 in one of the first
larger studies factors that correlate significantly with programming 
productivity (measured in effort per SLOC). Several of the later publications
use the same or a variant of these factors. A number of the described
factors obviously decreased in importance over the decades. For example, 
\emph{chief programmer team usage} is not a common practice today. Also
with the more and more standardised hardware, \emph{previous experience
with operational computer} does not seem to be a problem anymore.
Nevertheless, the majority of factors, such as \emph{user participation},
\emph{overall constraints on program design} or \emph{previous experience
with programming language} are still valid.

Albrecht then proposed his famous \emph{function points} \cite{albrecht79}.
In this study, he analysed factors like the used programming language and
the project size.

\SubSection{1980--1989}

Brooks \cite{brooks81}  uses factors from Walston and Felix
\cite{walston77} as basis in his study at IBM. He found especially the
effects of program complexity and structured programming to be important.



Jones started with \cite{jones86} a series of books about programming
productivity. He was one of first that analysed various productivity factors
over various domains and could provide industry averages. He focused his
measures strongly on LOC and FP.

DeMarco and Lister \cite{demarco87} then aimed in a completely different
direction from the LOC- and FP-centred research. They point out that ``The major
problems of our work are not so much \emph{technological} as
\emph{sociological} in nature.'' They consider turnover as one of the central
factors influencing productivity. 
They also mention the importance of a
proper work place with windows, natural light, quietness, etc. They substantiate
this by showing that a noisy workplace with a high probability leads to more
defects. 
The used
language, years of experience, number of defects and salary do not have an
significant effect on productivity in their opinion. 

They further claim that ``Quality, far beyond that
required by the end user, is a means to higher productivity.'' 
They then discuss work interruption as important issue and
introduce the \emph{E-Factor} as ratio of uninterrupted hours
and body-present hours as measure for this.

Finally, they list six factors that they called ``teamicide'', i.e., measures
that are the main obstacles in building (or growing) teams that partially
repeat earlier mentioned factors: defensive management, bureaucracy,
physical separation, fragmentation of people's time, quality reduction of the
product, phony deadlines, and clique control.
In summary, DeMarco and Lister provided in \cite{demarco87} the first and still
most comprehensive work on the soft factors influencing productivity in software
development.

The most famous model that involves productivity is COCOMO by
Boehm \cite{boehm88}. It is a cost-estimation model in which the productivity
of the developers obviously plays a decisive role. These factors have been
derived empirically from a large project database. The factors are discussed
in more detail in section \ref{sec:2000} with COCOMO II.


\SubSection{1990--1999}





The 90s showed, maybe as a result of DeMarco and Lister, a stronger
interest in soft factors.
Rasch studies in \cite{rasch91} the effect of factors such as team member
rotation, role ambiguity and role conflict on job satisfaction and actually
quantifies them based on a survey.




Lakhanpal \cite{lakhanpal93} concentrated on characteristics of groups
and their influence on productivity. The cohesiveness and capability had
the strongest influence in 31 development groups, experience had the
weakest influence.



Brodbeck describes in \cite{brodbeck94b} that in a
survey, the projects with a higher communication effort also were more
successful. Even the intensity of internal communication is positively
correlated with project success. This is in contrast to common software
engineering belief that high communication effort hampers productivity.


Wohlin and Ahlgren describe factors and their impact on time to market
in \cite{wohlin95}. They use 10 different factors in their study, mostly
factors that are covered by the publications discussed so far. They also
include product complexity, methods and tools and requirements stability that
could be considered technical factors. 



Blackburn, Scudder, and Van Wassenhove \cite{blackburn96} studied the factors
and methods that improved productivity in Western European companies. They
found \emph{project duration} and \emph{team size} to be significant.



Chatzoglou and  Macaulay \cite{chatzoglou97} interviewed participants of
over a hundred software projects about several factors and their influence
on productivity. They found that experience, knowledge and persistence of
the team members is considered important. Also the motivation of the users
and their communication with the rest of the team plays a role. Finally,
the available resources, tools and techniques used and the management style
are important factors.

Glass summarised in \cite{glass98} his findings on project ``runaways''. He
states that common causes for such failing projects are that they are huge,
that there are usually a multiplicity of causes and that they were aimed to
be ``breakthroughs'' in comparison to older systems. However, he also
suggests that technology is increasingly often the cause for project failure.

Hill et al.~\cite{hill98} investigated the influence of virtual offices
on aspects of work. Most interestingly in the productivity context is
that the perception that ``teamwork has been diminished''.

Port and McArthur \cite{port99} analysed the introduction of object-oriented
methods at Hughes Space and Communications. They found that an
object-oriented development approach coupled with object-oriented
implementation improves overall project productivity.


\SubSection{2000--2007}
\label{sec:2000}

The most thorough work in the area of productivity and its influencing
factors is COCOMO II by Boehm et al.~\cite{boehm00}. They have a long
experience in that area \cite{boehm88} and derive their factors from 
a large empirical body. Technical factors they identified are, for
example, precedentedness (how similar are the projects) or the product
complexity. Boehm et al. also analysed various soft factors 
and found that those factors combined are
more important than all the others. Those factors include
programmer capability and personnel continuity (turnover).

Jones states in \cite{jones00} that software projects are 
influenced by about 250 factors. Individual projects ``are usually 
affected by ten to 20 major
issues.'' 
Of course, he also investigated a series of soft factors. He 
lists and discusses
several factors based on case studies partially with quantitative results. 
36 of these factors are considered the major factors. In comparison to other
studies, he adds explicitly the support for modern telecommunication
facilities such as video conferencing.



Maxwell and Forselius argue in \cite{maxwell00} that the influencing factors
on productivity depend on the business domain the software is produced
for. For example, in the insurance domain \emph{requirements volatility},
\emph{software's logical complexity} and \emph{tools use} are significant
while in the public administration domain \emph{number of inquiries} and
\emph{customer participation} are of importance.





Kitchenham and Mendes \cite{kitchenham04} found that reuse is
taking place has a significant effect on productivity. The amount of reuse
is not that important. They also suggest that the productivity is not
significantly different in different countries.




\begin{table*}[htb]
\caption{The derived technical factors \label{tab:technical}}
\footnotesize
\begin{center}
\begin{tabular}{llr}

\hline 
\multicolumn{1}{c}{\textbf{\textsf{Factor}}} &
\multicolumn{1}{c}{\textbf{\textsf{Description}}} &
\multicolumn{1}{c}{\textbf{\textsf{No. of Sources}}} \\
\hline
\multicolumn{3}{c}{\textbf{Product}}\\
\hline
\textbf{Precedentedness} &
How similar are the projects? &
2\\
\textbf{Required Software Reliability} &
The level of reliability needed. &
3\\
\textbf{Database Size} &
How large is the data compared to the code? &
2\\
\textbf{Product Complexity} &
The complexity of the function and structure of the software. &
6\\
\textbf{Developed for Reusability} &
To what extent the components should be reusable. &
3\\
\textbf{Execution Time Constraints} &
How much of the available execution time is consumed. &
7\\
\textbf{Main Storage Constraint} &
How much of the available storage is consumed. &
3\\
\textbf{Software Size} &
The amount of code of the system. &
4\\
\textbf{Product Quality} &
The quality of the product influences motivation and hence productivity. &
2\\
\textbf{User Interface} & 
Degree of complexity of the user interface. &
3\\
\textbf{Development Flexibility} &
How strong are the constraints on the system? &
2\\
\textbf{Reuse} & 
The extent of reuse. &
2\\
\hline
\multicolumn{3}{c}{\textbf{Process}}\\
\hline
\textbf{Architecture Risk Resolution} &
How are the risks mitigated by architecture? &
1\\
\textbf{Process Maturity} &
The well-definedness of the process. &
1\\
\textbf{Platform Volatility} &
Time span between major changes. &
3\\
\textbf{Early Prototyping} &
Early in the process prototypes are built &
1\\
\textbf{Completeness of Design} &
The amount of the design that is completed when starting coding &
2\\
\textbf{Effective and Efficient V\&V} &
The degree to which defects are found and the needed effort. &
1\\
\textbf{Project Duration} &
Length of the project. &
2\\
\textbf{Hardware Concurrent Development} & 
Is the hardware developed concurrently? &
3\\
\hline
\multicolumn{3}{c}{\textbf{Development Environment}}\\
\hline
\textbf{Use of Software Tools} &
The degree of tool use. &
7\\
\textbf{Programming Language} &
The level of the used programming language. &
3\\
\textbf{Use of Modern Development Practices} &
Are modern methods applied? &
7\\
\textbf{Documentation match to life-cycle needs} &
How well the documentation fits to the needs. &
2\\

\hline
\end{tabular}
\end{center}
\end{table*}

Berntsson-Svensson and Aurum \cite{berntsson-svensson06} analysed in a survey
factors influencing project and product success. They found that different
industries define success in different terms. However, the identified
influencing factors are similar to other studies: well defined project scope,
complete and accurate requirements, good schedule estimations, customer/user
involvement, and adding extra personnel.




Mohagheghi and Conradi \cite{mohagheghi07} analysed especially the
connection between software reuse and productivity among other factors.
They show that there can be a strong positive influence.


Spiegl reports in \cite{spiegl07} on a survey on project management issues in
software development conducted with project managers. In terms of soft
factors he found that support of the top-management, business culture,
promotions, team building, relationship management and communication, freedom
and responsibility, and motivation and appreciation are important.

\Section{Results}
\label{sec:results}

As mentioned above, we roughly divide the productivity factors into 
\emph{technical} and \emph{soft} factors. We see soft factors as 
all non-technical factors influencing productivity. These factors mainly stem 
from the team and its work environment. Obviously, the borderline between
these two groups is sometimes blurry and is only intended to aid easier
comprehension. 

The technical factors are summarised in Table \ref{tab:technical}. In
this group we structured the factors in three categories. The \emph{product}
category contains all factors that have a direct relation to the product,
i.e., the system itself. The category \emph{process} is comprised of the
technical aspects of the process. Finally, the category \emph{development
environment} contains factors about the tools the developer uses in the
project. 

The soft factors are summarised in Table \ref{tab:soft}.
Overlapping factors are combined as far as possible. 
We employed a simple, non-unique 
categorisation to
aid a quick comprehension. \emph{Corporate Culture} contains the factors
that are on a more company-wide level whereas \emph{Team Culture} denotes
similar factors on a team level. In \emph{Capabilities and Experiences} are
factors summarised that are related to individuals. \emph{Environment} 
stands for properties of the working environment. Finally, project-specific
factors are in the \emph{Project} category.

\begin{table*}[htb]
\caption{The derived soft factors \label{tab:soft}}
\footnotesize
\begin{center}
\begin{tabular}{llr}

\hline 
\multicolumn{1}{c}{\textbf{\textsf{Factor}}} &
\multicolumn{1}{c}{\textbf{\textsf{Description}}} &
\multicolumn{1}{c}{\textbf{\textsf{No.~of Sources}}} \\
\hline

\multicolumn{3}{c}{\textbf{Corporate Culture}}\\
\hline
\textbf{Credibility} &
Open communication and competent organisation. &
4\\
\textbf{Respect} &
Opportunities and responsibilities. &
6 \\
\textbf{Fairness} &
Fairness in compensation and diversity. &
5\\
\hline
\multicolumn{3}{c}{\textbf{Team Culture}}\\
\hline
\textbf{Camaraderie} &
Social and friendly atmosphere in the team. &
1\\
\textbf{Team Identity} &
The common identity of the team members. &
2\\
\textbf{Sense of Eliteness} &
The feeling in the team that they are ``superior'' &
3\\
\textbf{Clear Goals} &
How clearly defined are the group goals? &
3\\
\textbf{Turnover} &
The amount of change in the personnel. &
7\\
\textbf{Team Cohesion} &
The cooperativeness of the stakeholders. &
9\\
\textbf{Communication} &
The degree and efficiency of which information flows in the team. &
4\\
\textbf{Support for Innovation} &
To what degree assistance for new ideas is available. &
1 \\
\hline
\multicolumn{3}{c}{\textbf{Capabilities and Experiences}}\\
\hline
\textbf{Developer Temperaments} &
The mix of different temperaments on the team. &
1\\
\textbf{Analyst Capability} &
The skills of the system analyst. &
8\\
\textbf{Programmer Capability} &
The skills of the programmer &
10\\
\textbf{Applications Experience} &
The familiarity with the application domain. &
7\\
\textbf{Platform Experience} &
The familiarity with the hard- and software platform. &
7\\
\textbf{Language and Tool Experience} &
The familiarity with the programming language and tools. &
8\\
\textbf{Manager Capability} &
The control of the manager over the project. &
7\\
\textbf{Manager Application Experience} &
The familiarity of the manager with the application. &
2\\
\hline
\multicolumn{3}{c}{\textbf{Environment}}\\
\hline
\textbf{Proper Workplace} &
The suitability of the workplace to do creative work. &
3\\
\textbf{E-Factor} &
This environmental factor describes the ratio of
uninterrupted hours and
body-present hours. &
2\\
\textbf{Time Fragmentation} &
The amount of necessary ``context switches'' of an employee. &
1\\
\textbf{Physical Separation} &
The team members are distributed over the building or multiple sites. &
4\\
\textbf{Telecommunication Facilities} &
Support for work at home, virtual teams, video conferencing with clients. &
2\\
\hline
\multicolumn{3}{c}{\textbf{Project}}\\
\hline
\textbf{Schedule} &
The appropriateness of the schedule for the development task. &
5\\
\textbf{Requirements Stability} &
The number of requirements changes. &
6\\
\textbf{Average Team Size} &
Number of people in the team. &
10\\
\hline
\end{tabular}
\end{center}
\end{table*}

In general, what is surprising in the studies is that communication effort
is positive for productivity. It is often discussed that communication
should be reduced to decrease ``unnecessary'' work. However, it seems the
problem is only that with increasing people the communication effort
increases strongly. Yet, a high fraction of effort on communication seems
like a good investment.

Then there is some agreement in the few studies that analysed these factors
that the business domain plays a role. Either the domain itself has an
influence on productivity or at least it determines which of the other
factors have the strongest influence. This contradicts general and generic
productivity models but suggests that individual models are needed.

It is also notable that although experience is often brought up and is
in interviews considered important, in empirical studies it is rather
insignificant. By far more interesting is the capability of the developers.
Hence, this suggests that only being in a profession for a long time
does not make one productive. 

\Section{Related work}
\label{sec:related}

An early review of the state of the art in software development productivity
was done by Jeffery and Lawrence \cite{jeffery81}. They concentrated on
the conflicting results w.r.t.\ some factors such as experience or size
that in some studies were found to have a positive in others a negative
effect. We do not consider that in our paper but only analyse the relevance
of a factor in general.

Maxwell, Van Wassenhove and Dutta \cite{maxwell96} relate their research
on productivity factors in military software projects to earlier studies
in other areas and the factors found there. however, this work is
already 12 years old and a large number of studies have contributed
to the knowledge about productivity factors since then.

Ram\'{i}rez and Nembhard \cite{ramirez04} analysed the more general
category of \emph{knowledge worker} productivity. Software developers
are in their work part of these knowledge workers (KW) as opposed to manual
workers. They state that ``it seems to be of common agreement that to date
there are no effective and practical methods to measure KW productivity.''
Hence, they concentrate on a review of the dimensions used in the literature
whereas our review considers the factors influencing productivity.

\Section{Conclusions}
\label{sec:conclusions}

The productivity of the development team is decisive for successful 
software projects. Hatton \cite{hatton07} shows that there are large
differences, especially in the abilities of the developers. ``[\ldots] in
most experiments, analysts regularly record variations of a factor of 10 or
more in the individuals' performance.'' This illustrates the large
potential for improvements in development projects.

However, controlling productivity is only possible if the influencing
factors are known. ``You cannot control what you cannot measure.''
\cite{demarco87}  Hence, a clear list of influences on productivity
in software development is needed in order to organise corresponding
analysis and control activities. Existing productivity models and
methods already make use of lists of productivity factors.

Yet, there is a large body of literature on productivity and productivity
factors accumulated over the last decades. This paper provides a systematic
review of this literature and a derived list of important factors based
on their use in the studies. Soft and technical factors are investigated
in equal detail and a list of factors is provided for each.

This list can now be used for modelling productivity and for productivity
improvement methods. For example, the \pf\footnote{ProdFLOW is a registred
trademark of the Siemens AG.} method described in
\cite{ruhe08} uses interview techniques for determining the most influential
factors in productivity for a specific organisation. These interviews can
be supported by the comprehensive knowledge about existing factors from the
compiled lists.

For further research, we need to add further detail to the lists of factors
by determining whether the factors influence productivity positively 
or negatively which is important for productivity models. Furthermore, there
are influences between factors that can also have significant effects that
need to be considered in this list and corresponding models.

Finally, for further surveys like this, it would be extremely useful if
the researchers that report about the influence of specific factors on
productivity were describing the factors, the measurment units and the
context in more detail. Then the knowledge can be aggregated in ways that
can provide even more value.

\balance
\bibliographystyle{latex8}
\bibliography{space08_survey}

\end{document}